\documentclass[
  superscriptaddress,
  reprint,
  aps,
  pra,
  amsmath,
  amssymb,
  floatfix,
  nofootinbib
]{revtex4-2}

\usepackage{amsmath}
\usepackage{amssymb}
\usepackage{amsfonts}
\usepackage{mathtools}
\usepackage{bm}
\usepackage{dsfont}
\usepackage{amsthm}
\usepackage{braket} 

\usepackage{graphicx}
\usepackage{booktabs}
\usepackage{array}
\usepackage{wrapfig}
\usepackage{float}

\usepackage{algorithm}
\usepackage{algpseudocode}
\usepackage[caption=false]{subfig}
\usepackage{caption}

\usepackage{ragged2e} 

\makeatletter
\AtBeginDocument{%
  \renewcommand\@makecaption[2]{%
    \par\addvspace\abovecaptionskip
    \begingroup
      \small\justifying
      \setlength{\parindent}{0pt}
      \noindent #1.\ #2\par
    \endgroup
    \addvspace\belowcaptionskip
  }%
}
\makeatother

\algrenewcommand\algorithmicrequire{\textbf{Input:}}
\algrenewcommand\algorithmicensure{\textbf{Output:}}

\usepackage[english]{babel}
\usepackage{CJKutf8} 
\usepackage{color}
\usepackage[normalem]{ulem}
\usepackage{url}
\usepackage{comment}

\usepackage{hyperref}
\hypersetup{colorlinks=true}

\newcommand{\needcite}[1]{\textcolor{red}{[Ref needed]}}

\newcommand{\Leiden}{%
  \affiliation{$\langle aQa^{L}\rangle$ Applied Quantum Algorithms, Leiden University, Netherlands}%
  \affiliation{LIACS, Leiden University, Netherlands}%
}
\newcommand{\ASU}{%
  \affiliation{School of Electrical, Computer, and Energy Engineering, Arizona State University, Tempe, Arizona 85281, USA}%
}

\begin{document}

\title{Riemannian gradient descent-based quantum algorithms for ground state preparation with guarantees}

\author{Mahum Pervez}
\email{mpervez@asu.edu}
\ASU

\author{Ariq Haqq}
\ASU

\author{Nathan A. McMahon}
\Leiden

\author{Christian Arenz}
\email{carenz1@asu.edu}
\ASU

\date{\today}

\begin{abstract}
We investigate Riemannian gradient flows for preparing ground states of a desired Hamiltonian on a quantum device. We show that the number of steps of the corresponding Riemannian gradient descent (RGD) algorithm that prepares a ground state to a given precision depends on the structure of the Hamiltonian. Specifically, we develop an upper bound for the number of RGD steps that depends on the spectral gap of the Hamiltonian, the overlap between ground and initial state, and the target precision. In numerical experiments we study examples where we observe for a 1D Ising chain with nearest-neighbor interactions that the RGD steps needed to prepare a ground state scales linearly with the number of spins. For all-to-all couplings a quadratic scaling is obtained. To achieve efficient implementations while keeping convergence guarantees, we develop RGD approximations by randomly projecting the Riemannian gradient into polynomial-sized subspaces. We find that the speed of convergence of the randomly projected RGD critically depends on the size of the subspace the gradient is projected into. Finally, we develop efficient quantum device implementations based on Trotterization and a quantum stochastic drift-inspired protocol. We implement the resulting quantum algorithms on IBM's quantum devices and provide data for small-scale problems.
\end{abstract}

\maketitle


\section{Introduction}
\label{sec:introduction}
The preparation of the ground state of a desired many-body Hamiltonian is an important task in quantum information science with a variety of applications. Finding the ground state energy of electronic structure type Hamiltonians in chemistry~\cite{google2020hartree,gresch2025guaranteed} and material science~\cite{clinton2024towards,yoshioka2025krylov}, as well as solving combinatorial optimization~\cite{pelofske2024short,sciorilli2025towards} and machine learning~ problems    \cite{dunjko2020non,cho2024machine,pan2023deep,chen2024empowering} can all be carried out by preparing the ground state of the corresponding Hamiltonian on a quantum device. In fact, many engineering and computer science optimization tasks, such as Max-Cut, graph partitioning, scheduling, and resource allocation, can be solved on quantum computers by mapping the related QUBO problem to finding the ground state of an Ising Hamiltonian \cite{lucas2014ising,farhi2014quantum,mohseni2022ising,jiang2023efficient,bybee2023efficient,zhang2024review,si2024energy,ushijima2017graph,pelofske2021decomposition,bozejko2025optimal,stollenwerk2019quantum,fu2025solving,ajagekar2022hybrid,ohyama2023resource,butt2024quantum}. 

Since the advent of noisy intermediate-scale quantum devices hybrid quantum-classical approaches have emerged as promising candidates for preparing (or approximating) ground states on near term quantum hardware. Variational Quantum Algorithms (VQAs)~\cite{cerezo2021variational} represent a general class of such hybrid methods that combine parameterized quantum circuits with classical optimizers to minimize an objective (cost) function. In this approach, the parameterized quantum circuit evaluates the cost function while the classical optimizer provides the parameter update. Among VQAs, the Variational Quantum Eigensolver (VQE) aims to prepare the ground state on a quantum device by minimizing the expectation value of the target Hamiltonian \cite{farhi2014quantum,kandala2017hardware,parrish2019quantum,liu2019variational,barison2022variational,fujii2022deep,wang2024entanglement,wiedmann2025convergence}. A prominent example of a VQE is the Quantum Approximate Optimization Algorithm (QAOA) that solves the combinatorial optimization problem MaxCut by minimizing the energy expectation of an Ising Hamiltonian defined on a graph \cite{farhi2014quantum}.  

However, the variational approach to ground state preparation faces several challenges: (i) parameters are typically optimized within a fixed quantum circuit or \emph{ansatz} for which it is a priori unclear whether the quantum circuit is expressive enough to be able to reach the ground \cite{mcclean2016theory, Sim2019Expressibility, holmes2022connecting}. (ii) since the underlying optimization landscape is typically nonconvex, the search for the global optimum can become trapped in suboptimal solutions \cite{bittel2021training, wiedmann2025convergence}. 

The first challenge (i) can be addressed by adaptive quantum algorithms \cite{grimsley2019adaptive,tang2021qubit, magann2022feedback, magann2023randomized,wiedmann2025convergence,malvetti2024randomized, x8g1-7h1k} that sequentially grow a quantum circuit based on feedback from measurement data. However, while in this way the circuit is automatically grown to become expressive enough, the adaptive growth can get stuck as often gradient measurements are used to inform the circuit growth, similar to challenge (ii).  This challenge has recently been addressed by observing that adaptive quantum algorithms for ground state problems can be understood as variants of Riemannian gradient flows \cite{absil2008optimization,helmke2012optimization}; a powerful optimization framework in which a cost function is directly optimized over unitary transformations, rather than parameters of a quantum circuit \cite{schulte2010gradient}. The discretized solution to the differential equation that defines the Riemannian gradient flow constitutes a Riemannian gradient descent (RGD) algorithm that provably converges to the ground state for almost all initial states \cite{panageas2019first,malvetti2024randomized}.  This favorable convergence behavior inspired various adaptive approaches that implement RGD on quantum hardware by approximating the Riemannian gradient through efficient gate implementations \cite{wiersema2023optimizing,magann2023randomized,malvetti2024randomized}. Very recently, it has been shown that minimizing the energy through RGD on quantum computers can in fact be understood as implementing quantum imaginary time evolution, which in turn can provide runtime estimates for solving ground state problems through RGD \cite{gluza2024double,mcmahon2025equating}. 

However, there is a key difficulty in using adaptive quantum circuits for implementing RGD. Either the circuit depth needed grows exponentially with the number of RGD steps \cite{gluza2024double} (or a circuit with exponential width is needed \cite{PhysRevLett.134.180602, alghadeer2025double}), or guarantees on finding the solution are lost when Riemannian gradients are approximated,  e.g. by projecting the RGD generator onto 1- and 2-local Paulis for efficient quantum device implementations \cite{wiersema2023optimizing}. 

In this work, we aim to overcome these challenges by developing randomized implementations of RGD. Introduced by some of the authors in \cite{magann2023randomized, malvetti2024randomized,mcmahon2025equating}, a randomized approach has the advantage that each step of RGD is efficiently implementable on a quantum device while convergence to the ground state can still be guaranteed almost surely \cite{malvetti2024randomized}. While in the works \cite{  magann2023randomized,malvetti2024randomized} only a single random direction was considered in each RGD step, making convergence slow, here we consider projecting the Riemannian gradient into polynomial sized subspaces to speed up convergence. We will build on the stochastic protocol introduced in \cite{mcmahon2025equating} to implement stochastic RGD on quantum device hardware.    

\begin{figure}[t!]
\centering
\includegraphics[width=0.95\columnwidth]{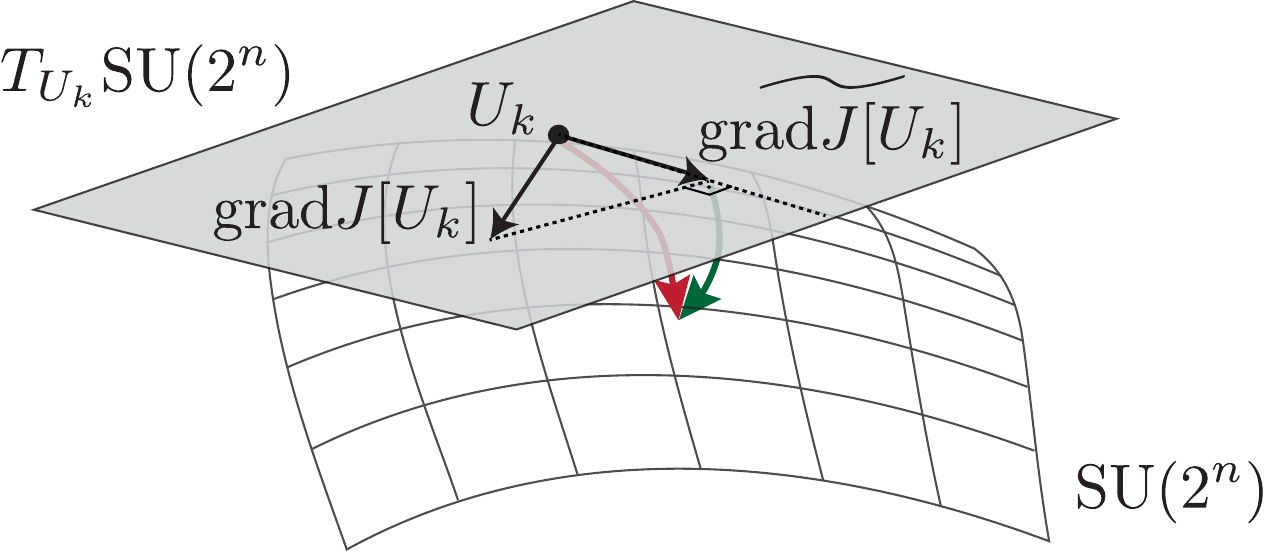}
\caption{\label{fig:introfig}
Schematic representation of the Riemannian gradient $\mathrm{grad}\,J[U_{k}]$ at a point $U_{k}$ in the tangent space $T_{U_{k}}\mathrm{SU}(2^{n})$ (illustrated as two-dimensional) for an $n$-qubit system. The Riemannian gradient is approximated by $\widetilde{\mathrm{grad}\,J[U_{k}]}$ that lies in a subspace (illustrated as one-dimensional) of dimension polynomial in $n$, enabling efficient implementation of the retraction $e^{\gamma\widetilde{\mathrm{grad}\,J[U_{k}]}}$ (red arrow) as an adaptive step in \eqref{eq:discretizedsolution} for ground-state preparation on a quantum device.
}

\end{figure}

The paper is organized as follows.
In Section~\ref{sec:theory} we start by presenting the theory of Riemannian gradient flows and its solution in discretized form, the Riemannian gradient descent for a cost function given by the expectation value of a Hamiltonian $H$. We go on to analyze the convergence behavior of RGD. We develop an upper bound for the number of steps required to prepare the ground state to a given precision through RGD and numerically analyze the convergence behavior for Ising Hamiltonians on a graph. We show that for a 1D Ising chain with nearest neighbor interactions a linear scaling with the number of spins $n$ is obtained while for a complete graph the number of steps required to prepare the ground state through RGD scales quadratically in $n$.  In  Section~\ref{sec:approximation} we introduce and study a stochastic variant of RGD that uses random projections into polynomially sized subspaces of the Riemannian gradient, as depicted in Fig. \ref{fig:introfig}. We compare its performance with the exact RGD method for different subspace dimensions. Finally, in Section~\ref{sec:implementation} we present a quantum device implementation of the randomized RGD algorithms. We compare implementations through Trotterization and a modified quantum stochastic drift protocol inspired by \cite{mcmahon2025equating}, and analyze their performance in terms of circuit depth and convergence behavior. Finally, we present quantum device data obtained from running the developed randomized algorithms on IBM's quantum devices for small scale problems.  

\section{Riemannian Gradient Flows for Ground-State Problems}
\label{sec:theory}
A Riemannian gradient flow on the unitary group is defined by the differential equation
\begin{align}
\label{eq:gradflowdef}
\frac{d}{dt}U(t) = \mathrm{grad}\, J[U(t)],
\end{align}
for a unitary operator $U(t) \in \mathrm{SU}(d)$ parameterized by some $t\in\mathbb R$, where $\mathrm{SU}(d)$ denotes the special unitary group of dimension $d$. Here, $\mathrm{grad}\, J[U] \in T_{U}\mathrm{SU}(d)$ is the Riemannian gradient of a cost function $J[U(t)]$ that lives in the tangent space $T_{U}\mathrm{SU}(d)$ which, at the identity, is $T_{\mathds{1}}\mathrm{SU}(d)=\mathfrak{su}(d)$, the Lie algebra $\mathfrak{su}(d)$ of traceless and skew-Hermitian matrices (see Fig. \ref{fig:introfig}). We consider a cost function given by the expectation value $J[U(t)] = \langle \psi(t) | H | \psi(t) \rangle$, where $|\psi(t)\rangle = U(t)|\psi(0)\rangle$ is the state obtained from an initial state $|\psi(0)\rangle$ and $H$ is the problem Hamiltonian whose ground state we aim to prepare.

\subsection{Riemannian Gradient Descent}
\label{sec:rgd}
The Riemannian gradient flow can be used to construct quantum algorithms that prepare the ground state of $H$. Consider the discretized solution
\begin{align}
\label{eq:discretizedsolution}
U_{k+1} = e^{\gamma \,\mathrm{grad}\,J[U_{k}]}\,U_{k}, \qquad U_{0}=\mathds{1},
\end{align}
to the differential equation in \eqref{eq:gradflowdef}, where $\gamma > 0$ is the step size. Here the Riemannian gradient of $J[U_{k}] = \langle \psi_{k} | H | \psi_{k} \rangle$ takes the form $\mathrm{grad}\,J[U_{k}] = [\,|\psi_{k}\rangle\langle \psi_{k}|,\, H\,] \in \mathfrak{su}(d)$, where $|\psi_{k+1}\rangle=e^{\gamma\text{grad}J[U_{k}]}\ket{\psi_{k}}$ is the state at step $k+1$ where $k=0,1,2,\cdots$ with $\ket{\psi_{0}}$ being the initial state \footnote{In a slight abuse of notation, the right multiplication by $U$ is omitted since it cancels in the solution \eqref{eq:discretizedsolution}}. The update \eqref{eq:discretizedsolution} defines a Riemannian gradient-descent (RGD) on $\mathrm{SU}(d)$. For sufficiently small $\gamma$ we have $J[U_{k+1}] \leq J[U_{k}]$, i.e., the expectation value of $H$ decreases monotonically as the circuit is grown according to \eqref{eq:discretizedsolution}.

\subsection{Convergence Behavior}
\label{sec:convergence}
The circuit growth in \eqref{eq:discretizedsolution} halts at a critical point $U^{*}$ where the Riemannian gradient vanishes, i.e., $\mathrm{grad}\,J[U^{*}]=0$. The critical-point structure has been analyzed in detail in ~\cite{rabitz2004quantum,hsieh2009topology}, showing that the critical points consist only of saddle points and global optima. Since the commutator $[\,|\psi_{k}\rangle\langle \psi_{k}|,\, H\,]$ vanishes only at eigenstates of $H$, unitaries that prepare eigenstates correspond to critical points of the Riemannian gradient. The ground states $|E_{0}\rangle$ and the highest excited states are global optima, whereas all other eigenstates are strict saddle points with at least one negative Hessian eigenvalue \cite{malvetti2024randomized,wiedmann2025convergence}. Because strict saddles are avoided with probability $1$ under first-order methods ~\cite{panageas2019first}, the RGD update \eqref{eq:discretizedsolution} converges to the ground state almost surely for almost all initializations ~\cite{malvetti2024randomized}, although approaching a saddle point can significantly slow down convergence \cite{zander2025role}.

It was recently shown \cite{gluza2024double,mcmahon2025equating} that the adaptive growth \eqref{eq:discretizedsolution} with an appropriately chosen step size $\gamma$ prepares the state
\begin{align}
\label{eq:ITE_State}
|\psi(\beta)\rangle = \frac{e^{-\beta H} \psi_{0}\rangle}{\|\,e^{-\beta H}|\psi_{0}\rangle\,\|},
\end{align}
generated by imaginary-time evolution $e^{-\beta H}$, where $\beta$ is the imaginary time and $\|\cdot\|$ denotes the Euclidean vector norm. Since $|\psi(\beta)\rangle \to |E_{0}\rangle$ as $\beta \to \infty$ provided the initial state has nonzero overlap $C_{0}=\langle E_{0}|\psi_{0}\rangle$, RGD \eqref{eq:discretizedsolution} converges to $|E_{0}\rangle$ whenever $C_{0}\neq 0$. In ~\cite{mcmahon2025equating} it was proven that if $\gamma=\beta/N$ for an integer $N$, then after $N$ steps the error $\epsilon=\|\,|\psi(\beta)\rangle-|\psi_{N}\rangle\,\|$ between the state in \eqref{eq:ITE_State} and the state $|\psi_{N}\rangle = e^{\frac{\beta}{N}\mathrm{grad}\,J[U_{N-1}]}|\psi_{N-1}\rangle$ generated by $N$ steps of RGD is upper-bounded by
\begin{align}
\label{eq:bound}
\epsilon \le \frac{5}{2}\,\frac{\beta}{N}\,\|H\|_{\infty}\big(e^{4\beta\|H\|_{\infty}}-1\big),
\end{align}
where $\|\cdot\|_{\infty}$ denotes the spectral norm. Using the triangle inequality, $\|\,|E_{0}\rangle-|\psi_{N}\rangle\,\| \le \|\,|E_{0}\rangle-|\psi(\beta)\rangle\,\| + \epsilon$, and the bound in \eqref{eq:bound}, the number of steps $N$ required to prepare the ground state through RGD to a desired precision $\varepsilon=\Vert |E_{0}\rangle-|\psi_{N}\rangle\ \Vert$ can be upper bounded by identifying $\beta$ for which $\Vert\ket{\psi(\beta)}-\ket{E_{0}}\Vert =\frac{1}{2}\varepsilon$. This is achieved by taking $\beta=\ln\!\left(\left(\frac{4}{|C_{0}|\varepsilon}\right)^{1/\Delta}\right)$, where $\Delta$ is the spectral gap of $H$, which yields 
\begin{align}
\label{eq:bound2}
N \le \frac{5}{2}\,\xi\,\frac{\ln(\alpha_{\varepsilon})}{\varepsilon}\,\alpha_{\varepsilon}^{\,4\xi}, 
\quad \alpha_{\varepsilon}=\frac{4}{|C_{0}|\varepsilon},\ \xi=\frac{\|H\|_{\infty}}{\Delta}.
\end{align} 
Thus, the complexity for preparing the ground state through RGD is at most exponential in $\frac{\Vert H\Vert_{\infty}}{\Delta }$ and $1/\text{poly} (\varepsilon |C_{0}|)$ in the initial state overlap $C_{0}$ with the ground state and the precision $\varepsilon$ to prepare the ground state. We remark here that a similar proof technique as in \cite{mcmahon2025equating} has been followed in \cite{robbiati2024double} to develop fidelity bounds for ground state preparation through RGD \cite{gluza2024double}.

The explicit dependence of the Riemannian gradient on the Hamiltonian $H$ suggests that the number of steps $N$ required to reach the ground state depends on the problem structure. In particular, the bound in \eqref{eq:bound2} ties the speed of convergence to spectral properties of $H$ and to the initial state overlap $C_{0}$.

As an example, consider Grover search on a $d=2^{n}$ dimensional Hilbert space, where $C_{0}=1/\sqrt{d}$ and $H=\mathds{1}-|\omega\rangle\langle\omega|$ formed by the projector $|\omega\rangle\langle\omega|,~\omega\in \{0,1\}^{n} $ that correspond to the solution of the search problem. In this case $\|H\|_{\infty}/\Delta = 1$, implying $N=\mathcal{O}(\sqrt{d}/\varepsilon)$, which corresponds to the optimal circuit complexity for unstructured search ~\cite{bennett1997strengths}. In fact, it has recently been shown that, Grover's algorithm can be interpreted as RGD that cools the system \cite{suzuki2025grover}.

More generally, when $\|H\|_{\infty}/\Delta=\mathcal{O}(1)$, the ground state can be prepared to precision $\varepsilon$ in $N = \mathrm{poly}^{-1}(\varepsilon |C_{0}|)$ steps of RGD.

\begin{figure}[h!]
\centering
\includegraphics[width=0.95\columnwidth]{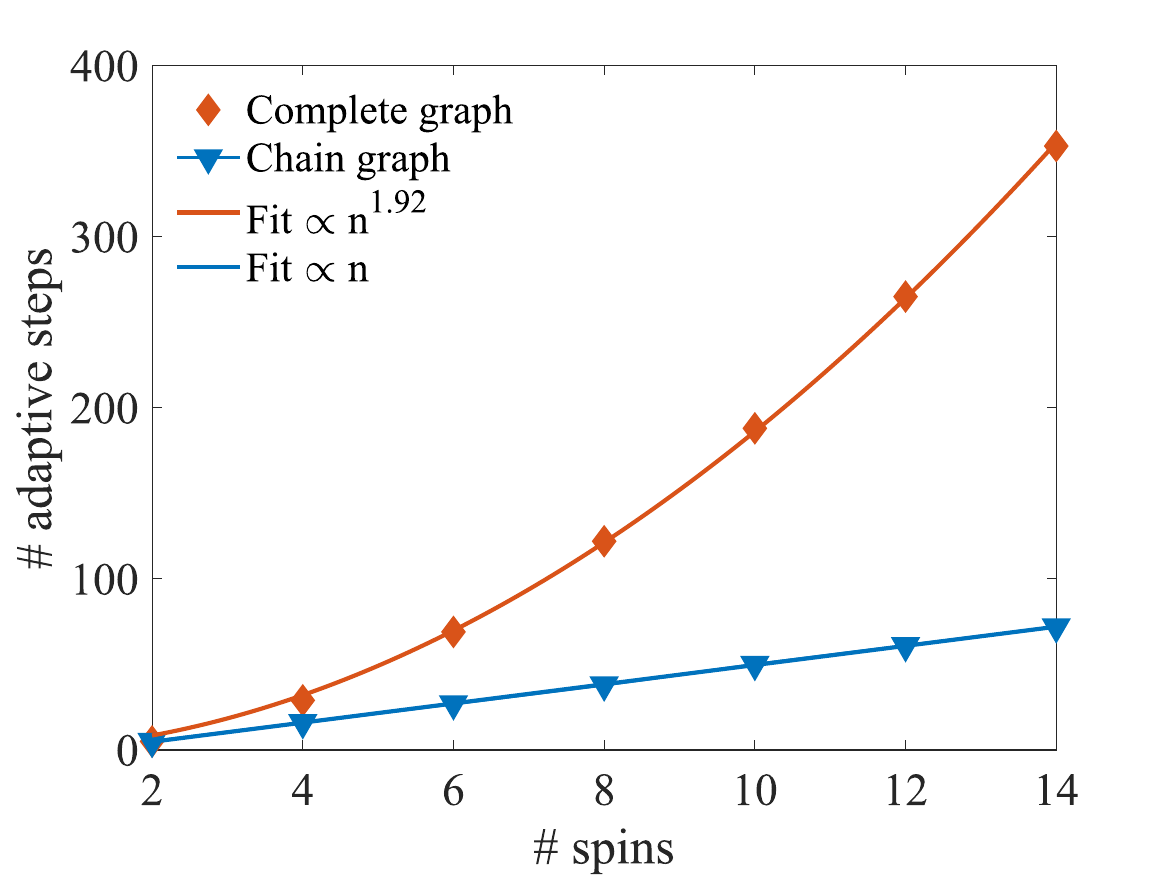}
\caption{\label{fig:ScalingGradFlow}
Number of adaptive steps required to prepare the ground state of the Ising Hamiltonian \eqref{eq:IsingHam} using Riemannian gradient descent defined by  \eqref{eq:discretizedsolution} for a 1D chain with nearest-neighbor interactions (blue) and a complete graph (red) versus the number of spins $n$. Solid lines indicate an approximately linear fit (chain) and a quadratic fit (complete graph), respectively.
}
\end{figure}

In Fig.~\ref{fig:ScalingGradFlow} we numerically investigate the number of steps needed to prepare the ground state of the Ising Hamiltonian
\begin{align}
\label{eq:IsingHam}
H=\sum_{(i,j)\in E} Z_{i}Z_{j},
\end{align}
defined on a graph $G(V,E)$, where $V$ denotes the set of vertices given by a set of $n$ spins, $E$ is the set of edges describing the interactions between the spins. We assume equal edge weights and $Z_{i}$ is the Paul-Z operator acting non-trivially only on the ith spin.

We study in Fig. \ref{fig:ScalingGradFlow} the scaling of the number of adaptive steps needed to prepare the ground state versus the number of spins $n$ in a graph for a chain graph (blue) and a complete graph (red). In both cases, the initial state is an equal superposition state $|\psi_{0}\rangle = |+\rangle^{\otimes n}$, and the step size is fixed as
\begin{align}
\label{eq:step_size}
\gamma = \frac{1}{4 \,\|H\|_{\infty}},
\end{align}
to ensure monotonic convergence ~\cite{magann2023randomized}. Data points correspond to reaching an approximation ratio of $0.99$ (defined by the ratio of the cost function value $J[U_{k}]$ and the ground state energy). 

We observe that the number of steps required to prepare the ground state depends critically on the graph structure: the chain exhibits approximately linear scaling, whereas the complete graph shows approximately quadratic scaling in $n$. This indicates that the speed of convergences of RGD contains information about the structure and empirical hardness of the ground state problem at hand.  We remark here that the bound in \eqref{eq:bound2} would yield an exponential scaling regardless of the type of graph considered. Developing tighter bounds that are able to capture the scaling observed in Fig. \ref{fig:ScalingGradFlow} will be subject of future studies. 

We further note that the associated Max-Cut problem can be solved in linear and quadratic time, respectively through a solely classical approach \cite{bondy1976graph, west2001introduction}.

\section{Approximating the Riemannian Gradient}
\label{sec:approximation}
To implement the RGD update step \eqref{eq:discretizedsolution} on a quantum computer, the adaptive growth $e^{\gamma\text{grad} J[U_{k}]}$ has to be synthesized through quantum logic gates. Since $\text{grad} J\in \mathfrak{su}(2^{n})$ lives in general in an exponentially large Lie algebra, i.e., $\text{dim}(\mathfrak{su}(2^{n}))=2^{2n}-1$ for a $n$-qubit system, the exact RGD update is generally not efficiently implementable. 

To address the challenge of efficiently implementing the update step $e^{\gamma\text{grad}J[U_{k}]}$ on a quantum device, several approaches \cite{wiersema2023optimizing,magann2023randomized,malvetti2024randomized,mcmahon2025equating} have been proposed that approximate the Riemannian gradient by considering projections of $\text{grad}J$ into polynomially sized subspaces $\mathcal A_{k}\subset \mathfrak{su}(2^{n})$. This is achieved by picking polynomially many (normalized) Pauli operators $iP_{j}$ that span $\mathcal A_{k}$ in each step $k$ to obtain an approximate gradient 
\begin{align}
\label{eq:approxGrad}
\widetilde{\text{grad}J[U_{k}]}=\sum_{iP\in\mathcal A_{k}}\langle \text{grad}J[U_{k}], iP\rangle iP,
\end{align}
where $\langle \cdot,\cdot\rangle$ denotes the Hilbert-Schmidt inner product. Since 
\begin{align}
\left .\frac{d}{d\theta}\bra{\psi_{k}}e^{i\theta P_{j}}He^{-i\theta P_{j}}\ket{\psi_{k}}\right |_{\theta=0}=\langle \text{grad} J,iP_{j}\rangle,  
\end{align}
the corresponding coefficients $\langle \text{grad} J,iP_{j}\rangle$ can be estimated through gradient measurements obtained through e.g., the parameter shift rule or finite differences ~\cite{mitarai2018quantum, schuld2019evaluating}. The unitary $e^{\gamma \widetilde{\text{grad}}J}$ generated by the approximated gradient $\widetilde{\text{grad}}J$ can then efficiently be  implemented through Trotterization \cite{suzuki1985decomposition}. However, while such approximations allow for efficient implementations of $e^{\gamma \widetilde{\text{grad}}J}$, convergence guarantees to the ground state are typically lost. 

This problem has been addressed in \cite{magann2023randomized} by utilizing \emph{random} projections in each adaptive step that are for example created through conjugating a Pauli operator by unitary 2-designs. This randomized approach ensures that the adaptive step is efficiently implementable while simultaneously ensuring convergence to the ground state almost surely \cite{malvetti2024randomized}. Since in this case the Riemannian gradient is only approximated through projecting onto a single random direction, the overall speed of convergence is slow, resulting in an $\mathcal O(2^{n})$ scaling for the number of adaptive steps needed to prepare the ground state.

\begin{figure}[!t]
\centering
\includegraphics[width=\columnwidth]{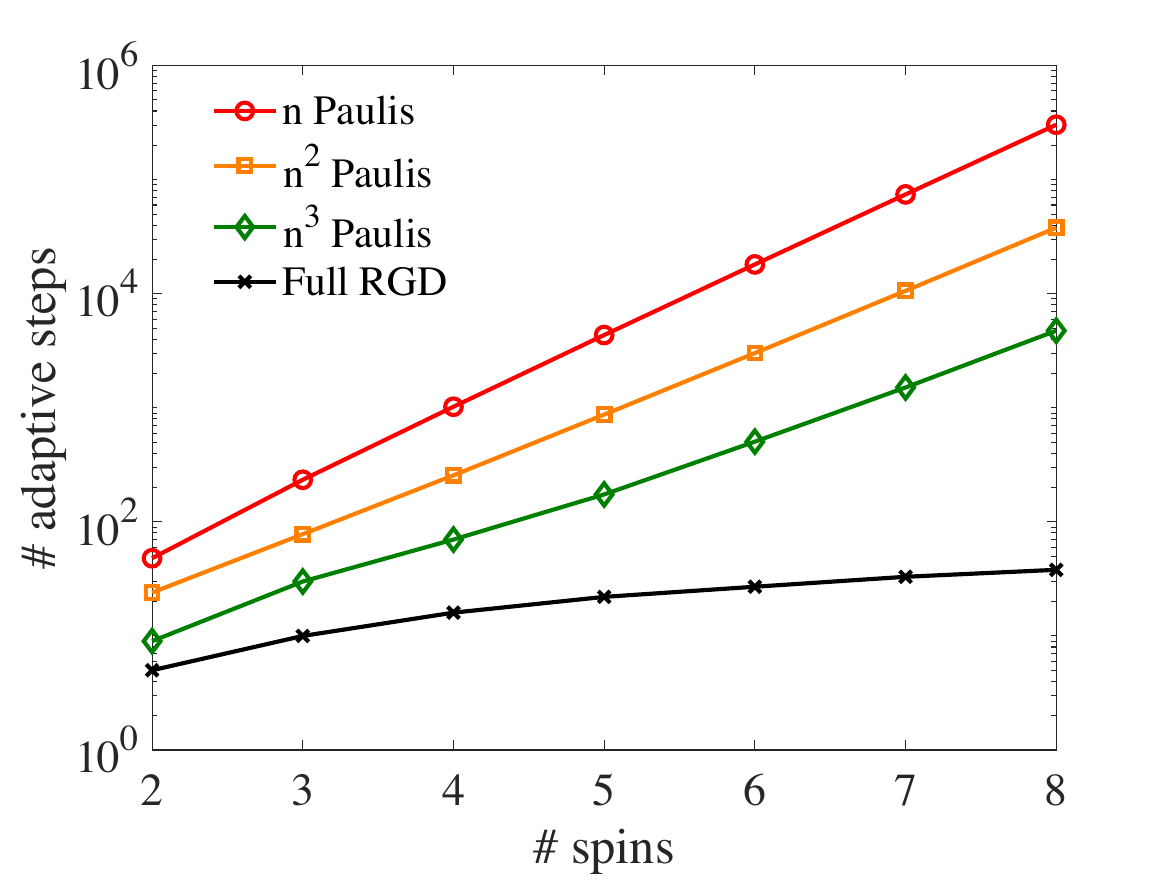}
\caption{\label{fig:ScalingApproxGradFlow} Number of a adaptive steps needed to prepare the ground state of the 1D Ising chain with nearest neighbor interactions \eqref{eq:IsingHam} for different approximations of the Riemannian gradient given by \eqref{eq:approxGrad} as a function of the number of spins $n$ on a semi-logarithmic scale. The red curve corresponds to randomly picking $n$ Pauli operators in each adaptive step and projecting the Riemannian gradient into the corresponding subspace.  The yellow and green curve was obtained by projecting into $n^{2}$ and $n^{3}$ dimensional subspaces. For comparison we show again the linear scaling behavior of the full RGD algorithm (blue curve in Fig. \ref{fig:ScalingGradFlow}) as a black curve. Each data point corresponds to an average taken over 100 samples.}
\end{figure}

In order to address this challenge, here we investigate \emph{random} projections into polynomial sized subspaces $\mathcal A_{k}$ in each adaptive step $k$. This is achieved by uniformly picking in each step $k$ a random set of Pauli operators $\{iP\}$ that span $\mathcal A_{k}$. The corresponding pseudo-code of this algorithm is shown in Algorithm 1. 

\begin{figure}[t]
\captionsetup{type=algorithm}
\caption{\textsc{Randomized-Subspace Riemannian Gradient Descent}}
\label{alg:approximate_rgf}

\centering
\begin{minipage}{\columnwidth}
\begin{algorithmic}[1]
\Require Number of qubits $n$; Hamiltonian $H$; initial state $\ket{\psi_0}$; number of adaptive iterations $N$; number of Trotter/qDRIFT steps $M$; step size $\gamma=\frac{1}{4 \Vert H \Vert_{\infty}}$; method $\in \{\text{Trotter}, \text{qDRIFT}\}$
\Ensure State $\ket{\psi}$ approximating the ground state of $H$ after $N$ steps
\State $\mathcal{P} \gets \{P\in\{I,X,Y,Z\}^{\otimes n} : P\neq I^{\otimes n}\}$
\Statex $\triangleright$ Full non-identity Pauli string set
\State $\ket{\psi} \gets \ket{\psi_0}$ 
\For{$k = 1$ to $N$}
    \State \textbf{Sample Pauli subspace:} choose $\mathcal{A}_k \subset i\mathcal{P}$ uniformly at random \emph{without} replacement
    \State \textbf{Approximate Riemannian gradient on $\mathcal{A}_k$:}
    \For{$iP \in \mathcal{A}_k$}
        \State \resizebox{0.87\linewidth}{!}{$
        g_{P}=\langle \mathrm{grad}\,J[U_k], iP \rangle 
        \gets 
        \left. \frac{d}{d\theta} 
        \bra{\psi} e^{i\theta P} H e^{-i\theta P} \ket{\psi} \right|_{\theta=0}$} 
         \Comment{Gradient component along $iP$}
    \EndFor

    \If{$\text{method} = \text{Trotter}$}
        \State \textbf{First-Order Trotter}
        \State $U \gets I$
        \For{$m=1$ to $M$}
            \ForAll{$iP \in \mathcal{A}_k$}
                \State $U \gets \exp\big(\gamma\, g_{P}\, iP/M\big)\, U$  \EndFor     
        \EndFor
    \ElsIf{$\text{method} = \text{qDRIFT-inspired}$}
        \State \textbf{qDRIFT}
        \State $\begin{aligned}[t]
        U 
        \gets \textsc{qDRIFT-inspired}\big(\mathcal{A}_k,g_{P},\gamma,M\big)
        \end{aligned}$

    \EndIf

    \State $\ket{\psi} \gets U \ket{\psi}$ 
\EndFor

\State \Return $\ket{\psi}$
\end{algorithmic}
\end{minipage}
\end{figure}

In Fig. \ref{fig:ScalingApproxGradFlow} we show the scaling of the number of adaptive steps needed to prepare the ground state of the Ising chain Hamiltonian with nearest neighbor interactions. We considered an approximated Riemannian gradient that is projected into a randomly selected subspace $\mathcal{A}_{k}$ of dimension linear in the number of spins (red), quadratic (yellow), and cubic (green). For comparison we show again the (linear) scaling of the full RGD as a black line. The step size $\gamma$ was again chosen according to \eqref{eq:step_size}, for the approximate Riemannian gradient.  

We note that even when $\mathrm{grad}\,J$ is approximated by
projections onto polynomially-sized subspaces, the worst-case scaling remains exponential in the system size. \emph{However}, by changing the size of the subspace we can reduce the number of adaptive steps required for convergence. While the circuit depth increases relative to using a single random projection when $e^{\gamma \widetilde{\text{grad}J}}$ is implemented through Trotterization, additionally randomizing the implementation of each gate $e^{i\gamma\langle \text{grad}J[U_{k}], iP\rangle P}$ allows for reducing the circuit depth to implement $e^{\gamma\widetilde{\text{grad}J}}$, which we study next.

\section{Quantum Device Implementations}
\label{sec:implementation}

In order to implement RGD on current quantum hardware we need to be able to simulate $e^{\gamma\widetilde{\text{grad}J[U_{k}]}}$ per adaptive step on a quantum device efficiently. This follows the standard problem of Hamiltonian simulation for which there are numerous approaches \cite{Lloyd1996,suzuki1985decomposition,campbell2019random, Berry2015Taylor,LowChuang2019,Childs2019PF}. The simplest approach would be to employ Trotterization \cite{suzuki1985decomposition} that approximates each approximated RGD update step according to 
\begin{equation}
e^{\gamma\,\widetilde{\text{grad}J[U_{k}]}}\approx \left(\prod_{iP\in\mathcal{A}_k} e^{i\gamma\langle \text{grad}J[U_{k}], iP\rangle P / M}\right)^M.
\end{equation}
Since typically the RGD step size $\gamma \ll 1$ is small to ensure monotonic convergence of RGD, we often can pick $M=1$ without suffering from a large approximation error. We thus have that 
\begin{align}
\label{eq:trottersim}
e^{\gamma\,\widetilde{\text{grad}J[U_{k}]}}\approx \prod_{iP\in\mathcal{A}_k} e^{i\gamma\langle \text{grad}J[U_{k}], iP\rangle P},
\end{align}
is often a good approximation of each approximated RGD update step. 
 However, this approach becomes infeasible on current quantum devices, particularly when $\widetilde{\text{grad}\,J[U_k]}$ is non-sparse as Trotterization applies all Pauli terms in the adaptive set $\mathcal{A}_k$, including those with negligible gradient contributions, thus increasing circuit depth significantly for minimal gain in gradient descent performance. To overcome this challenge we propose an alternate randomized method to simulate $e^{\gamma\widetilde{\text{grad}J[U_{k}]}}$. 

To further reduce the gate count, we adopt a quantum stochastic drift (qDRIFT) inspired  protocol~\cite{campbell2019random} that implements for sufficiently small steps sizes $\gamma$ the RGD update on average \cite{mcmahon2025equating}. 
For a generic step size, we pick a sufficiently large  ``qDRIFT'' repetition rate $M$ of Pauli product operators drawn from $\mathcal{A}_k$ with probabilities  
$\{ |\langle \mathrm{grad}J[U_k], iP\rangle|/\lambda_k \mid iP\in\mathcal{A}_k \}$, such that on average the unitary
\begin{align}
\label{eq:qdriftsim}
\prod_{j=1}^{M}
e^{\,i\,\lambda_k\,\gamma\, s_{k,j}/M \; P_j },
\end{align}
approximates $e^{\gamma\widetilde{\text{grad} J}}$ where
$$\lambda_k = \sum_{iP\in\mathcal{A}_k} 
\left| \langle \mathrm{grad}J[U_k], iP\rangle \right|,
$$
and
$$s_{k,j} = \mathrm{sign}\!\left(\langle \mathrm{grad}J[U_k], iP_j\rangle\right),
$$
contains the sign of the gradient, which would otherwise be lost under randomized sampling. 
The corresponding pseudocode of this algorithm is shown below in Algorithm 2. We note that throughout this work we use $M=1$ qDRIFT repetitions.  

\begin{figure*}[t]
\captionsetup[subfloat]{%
  position=top,
  justification=raggedright,
  singlelinecheck=false
}

\centering
\subfloat[\label{subfig:scaling_k-bodies}]{%
  \includegraphics[width=0.48\textwidth]{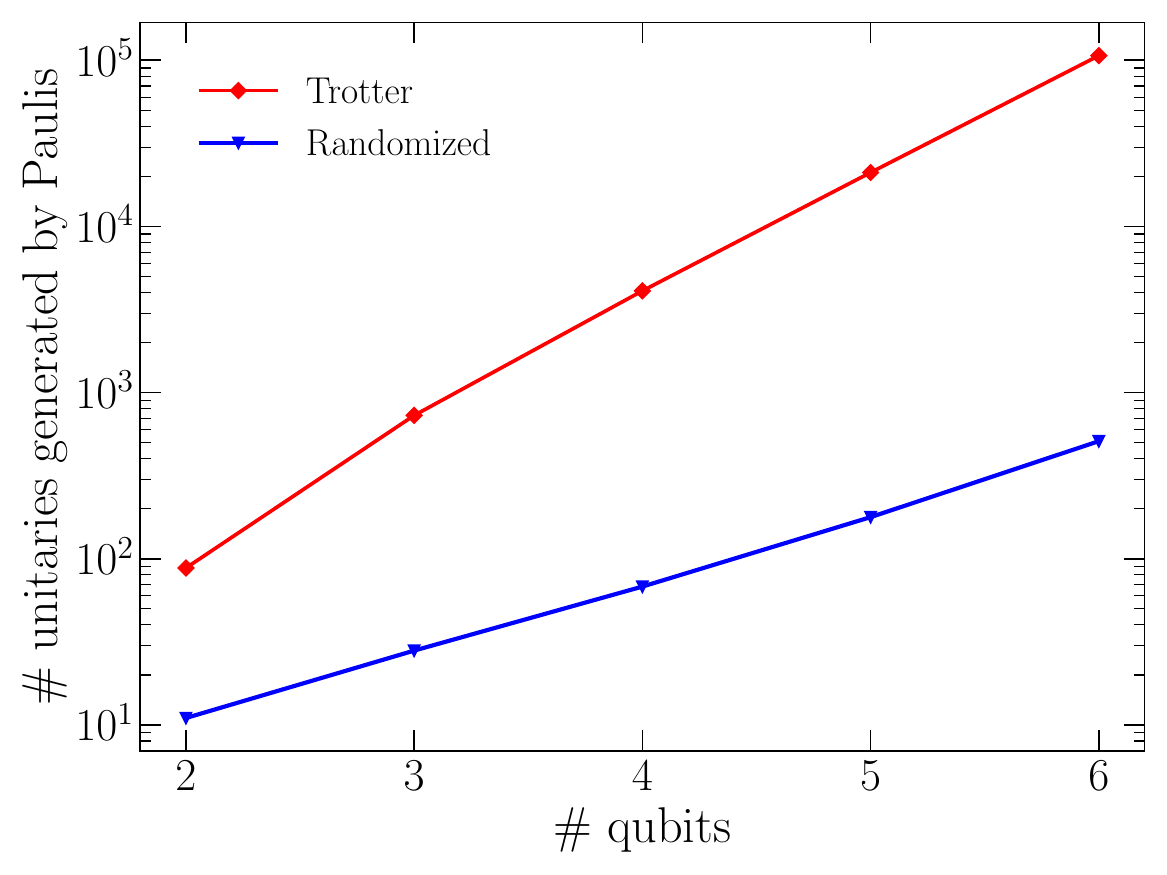}
}\hfill
\subfloat[\label{subfig:scaling_circuit_depth}]{%
  \includegraphics[width=0.48\textwidth]{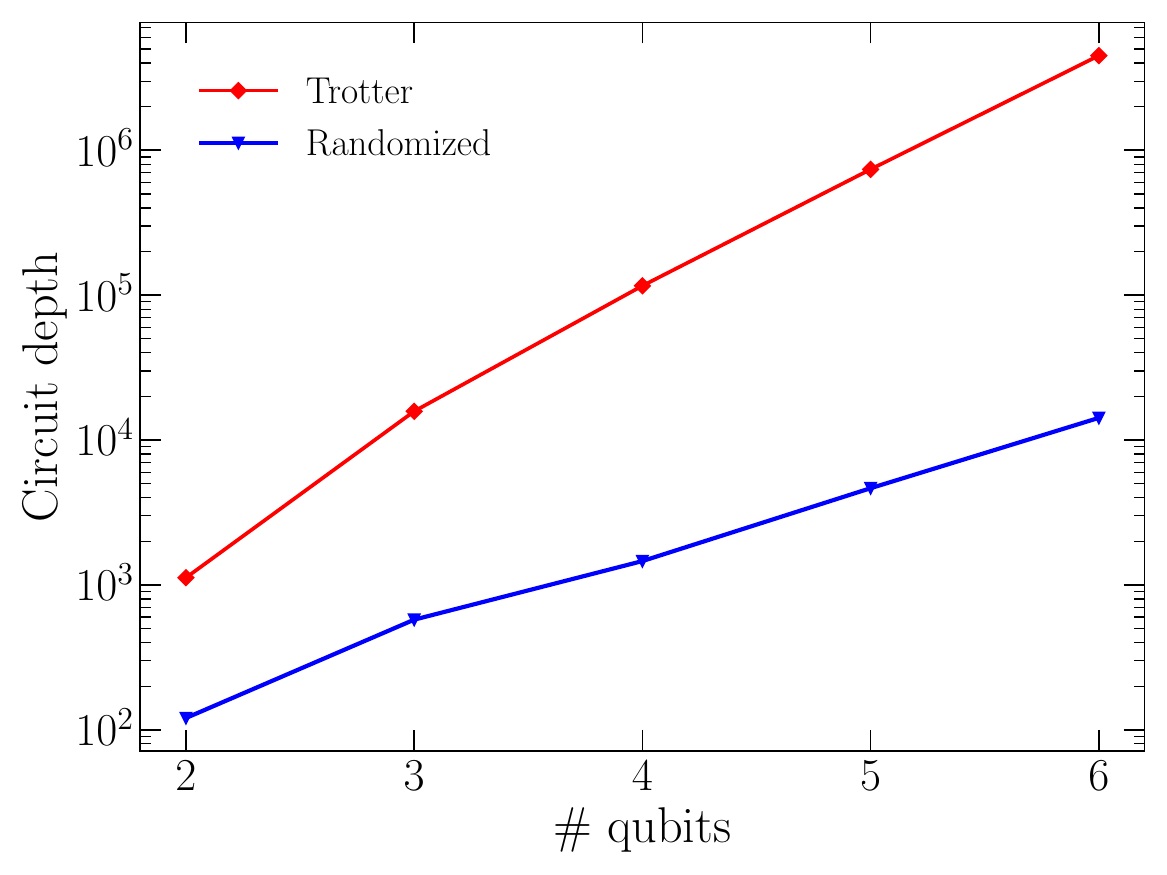}
}
\caption{Performance comparison of Trotter-based \eqref{eq:trottersim} and randomized (qDRIFT-inspired) implementations \eqref{eq:qdriftsim} for simulating the approximated RGD algorithm for preparing the ground state with an approximation ratio of $0.99$ of a 1D Ising Hamiltonian with nearest neighbor interactions. In both cases the Riemannian gradient is projected onto an $n^{3}$-dimensional subspace. A single Trotter step and qDRIFT repetition is used in all cases. (a) total number of unitaries generated by Pauli strings and (b) upper bound for the circuit depth as a function of the number of qubits/spins required to prepare the ground state.}
\label{fig:scaling_plots}
\end{figure*}

As shown in Fig.~\ref{fig:scaling_plots}, we evaluate two metrics as a function of the number of qubits:  
\begin{itemize}
    \item [(i)] the number of unitaries $e^{i\lambda\gamma s_{j}P_{j}}$ generated by Pauli strings $P_{j}$ required to reach an approximation ratio of $0.99$ [Fig.~\ref{subfig:scaling_k-bodies}].
    \item [(ii)] an upper bound of the circuit depth of the compiled circuit obtained by transpiling every possible Pauli rotation gate on IBM hardware, followed by summing the corresponding circuit depth [Fig.~\ref{subfig:scaling_circuit_depth}].    
\end{itemize}
In the numerical simulations the approximate Riemannian gradient is formed by projecting onto an $n^{3}$-dimensional subspace. First-order Trotterization implements $e^{\gamma \widetilde{\mathrm{grad}J[U_k]}}$ using a single Trotter step per iteration, whereas the randomized (qDRIFT-inspired) method picks one sample and applies \eqref{eq:qdriftsim}. The randomized approach is ran in its entirety $100$ times, then the resulting energy expectation is averaged.

\begin{figure}[h]
\captionsetup{type=algorithm}
\caption{\textsc{qDRIFT-inspired}: Randomized implementation of the approximated Riemannian Gradient Descent}
\label{alg:simulate_aprx}
\centering
\begin{minipage}{\columnwidth}
\begin{algorithmic}[1]
\Require Subspace $\mathcal{A}_k$; coefficients $\{g_P\}_{iP\in\mathcal{A}_k}$ with $g_P = \langle \mathrm{grad}\,J[U_k], iP\rangle$; step size $\gamma$; qDRIFT repetitions $M$
\Ensure Unitary $U$ implementing the qDRIFT product
\State $\lambda \gets \sum_{P \in \mathcal{A}_k} |g_P|$
\For{$j=1$ to $M$}
    \State Sample $M$ samples of $iP_j \in \mathcal{A}_k$ with probability  $|g_{P}|/\lambda$ \emph{with} replacement
    \State $s_j \gets \mathrm{sign}(g_{P_j})$
    \State $U \gets e^{\,i \lambda \gamma s_j P_j / M}\, U$
\EndFor
\State \Return $U$
\end{algorithmic}
\end{minipage}
\end{figure}

Figure~\ref{subfig:scaling_k-bodies} shows that both techniques exhibit similar asymptotic scaling with qubit count, but the randomized approach uses substantially fewer Pauli strings to achieve the same approximation ratio. Figure~\ref{subfig:scaling_circuit_depth} shows that this reduction directly translates into a lower compiled circuit depth—the randomized implementation consistently produces shallower circuits than first-order Trotterization, with the advantage increasing as the system size grows.

\begin{figure*}[t]
\captionsetup[subfloat]{%
  position=top,
  justification=raggedright,
  singlelinecheck=false
}

\centering
\subfloat[\label{subfig:2_qubit}]{%
  \includegraphics[width=0.48\textwidth]{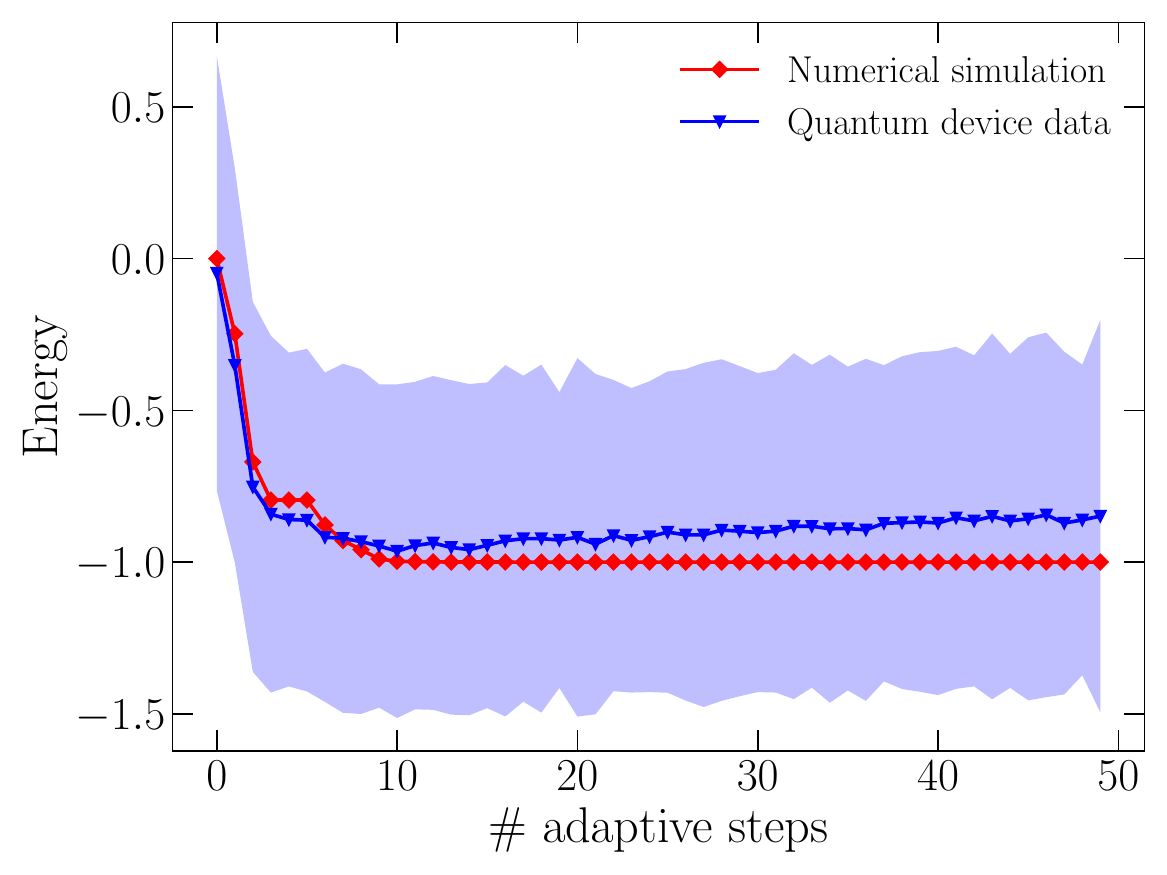}
}\hfill
\subfloat[\label{subfig:3_qubit}]{%
  \includegraphics[width=0.48\textwidth]{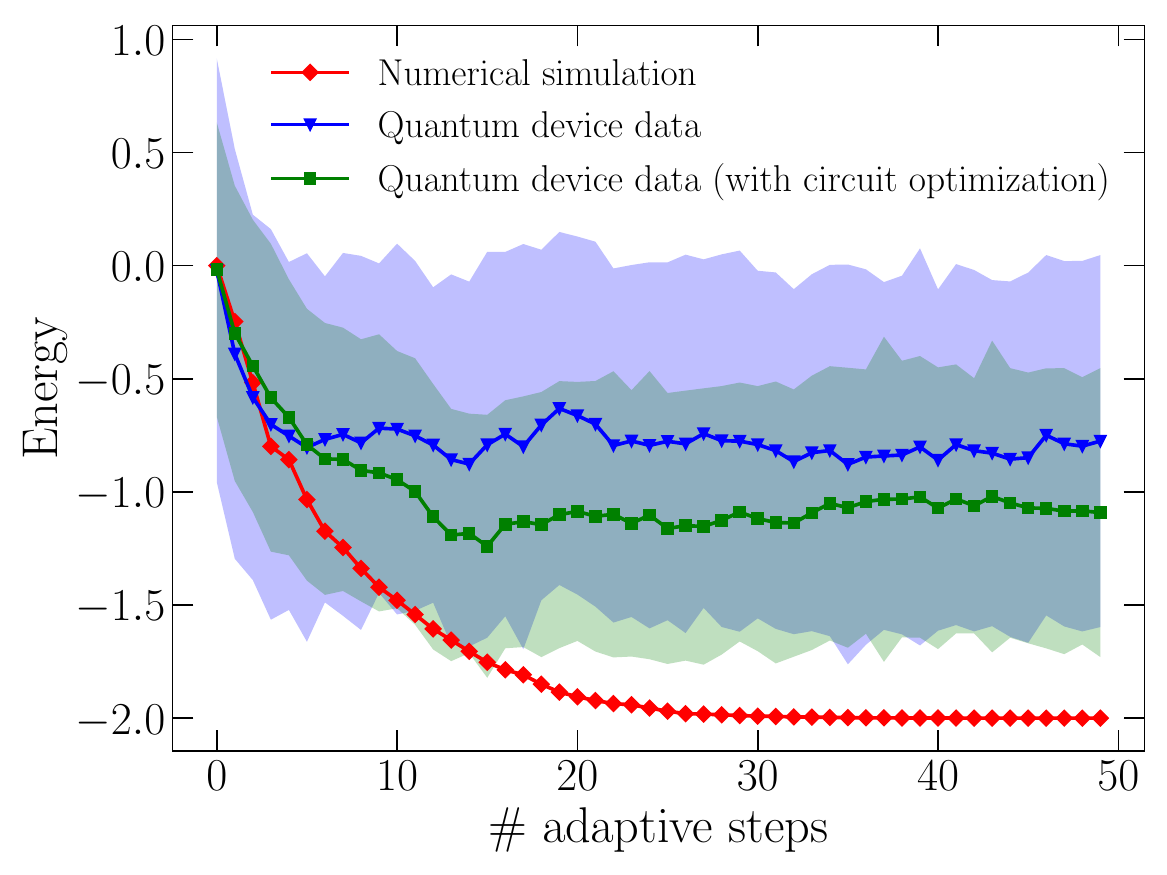}
}
\caption{Quantum device implementation 
of the randomized approximate RGD algorithms (Algorithm 1 + Algorithm 2) on IBM Quantum hardware (\texttt{ibm\_torino}) for a two qubit system (a) and a three qubit system (b). In both cases the average energy (estimated cost function value) is shown as a function of the number of adaptive steps for a a 1D Ising Hamiltonian with nearest neighbor interactions. The energy expectation reported is averaged over each iteration over $5$ runs and for each run the number of adaptive steps taken is $50$. The red curve shows for comparison a numerical simulation of the algorithm, the blue curve shows hardware results without circuit optimization, and the green curve shows hardware results with circuit optimization. The shaded region shows one standard deviation. }
\label{fig:qdrift_experimental}
\end{figure*}

Figure~\ref{fig:qdrift_experimental} shows experimental data obtained from implementing  the randomized approximate RGD algorithm (i.e, Algorithm 1 + Algorithm 2) on IBM hardware. In Fig.~\ref{subfig:2_qubit}, we run the algorithm on $n=2$ qubits with an Ising  Hamiltonian \eqref{eq:IsingHam} on a chain graph, projecting the Riemannian gradient onto an $n^{3}$-dimensional Pauli subspace. The energy expectation reported is averaged over each iteration over five runs and for each run the number of adaptive steps was chosen to be $50$. The ideal, noiseless evolution is shown in red, while the hardware execution without circuit optimization is shown in blue, with shaded regions indicating one standard deviation obtained over three repetitions. 

The quantum device results track the simulated trajectory closely and successfully converge to the expected ground state.  In Fig.~\ref{subfig:3_qubit}, we repeat the experiment for $n=3$ qubits. As system size increases, noise accumulation becomes more pronounced: the unoptimized hardware execution (blue) deviates significantly from the ideal curve, and even enabling circuit optimization (green) is insufficient to recover convergence to the ground state. These results highlight the impact of hardware noise and emphasize the need for low-depth circuit realizations when scaling beyond two qubits.

\section{Conclusion}
\label{sec:conclusion}
In this work we have developed and analyzed randomized  Riemannian gradient descent algorithms for ground state preparation on near-term quantum devices. We first showed that Riemannian gradient descent (RGD) exhibits problem-dependent convergence behavior: the number of RGD steps required to prepare the ground state to a desired precision scales linearly for a 1D Ising Hamiltonian with nearest neighbor interactions and quadratically for a Ising Hamiltonian with all-to-all coupling, i.e., defined on a  complete graph with equal edge weights. This convergence behavior suggests that RGD is not problem agnostic but can capture the structural properties of the underlying Hamiltonian and therefore can reflect the empirical hardness of the corresponding ground-state problem. 

We went on to show that approximating the Riemannian gradient through random projections into polynomially sized subspaces spanned by Pauli operators enables efficient implementations of each step of RGD on quantum computers while preserving convergence guarantees. We found that increasing the dimension of these random subspaces accelerates convergence and more closely tracks the behavior of the full Riemannian gradient descent, though at the cost of a higher circuit complexity. This trade-off highlights that subspace size can serve as a tunable parameter balancing convergence speed and circuit complexity.

We also developed and bench-marked two quantum device implementations based on (first-order) Trotterization and a quantum stochastic drift (qDRIFT) inspired protocol. We showed that both methods exhibit comparable asymptotic scaling, but the randomized implementation achieves significantly lower circuit depth. The reduction in Pauli terms directly translates into reduced compiled circuit depth, which is essential for mitigating decoherence on current quantum hardware. The quantum device implementation of the randomized RGD for a two-qubit system matches closely the numerical simulations. However, for three qubits, accumulated noise prevents convergence to a ground state even with transpiler-level circuit optimization. These results highlight both the capability and the present limits of implementing RGD-based ground-state preparation on current quantum hardware.

Overall, this work provides a foundation for Riemannian gradient-based quantum optimization in which each adaptive step can be implemented with polynomial resources, while still maintaining the global convergence guarantees of Riemannian gradient flows. Although full convergence remains exponentially costly in the worst case, the proposed randomized approach makes each individual step practical for near-term hardware. Continued development of such randomized manifold-based techniques, together with advances in quantum devices and compilation strategies \cite{PhysRevLett.134.180602, alghadeer2025double,xiaoyue2024strategies}, may further expand the feasibility of ground-state preparation for increasingly large systems, and of other optimization tasks \cite{robbiati2024double} and applications in quantum information science \cite{suzuki2025double}.~\\

\begin{acknowledgments}
N.A.M. was co-funded by the European Union (ERC CoG, BeMAIQuantum, 101124342). M. P. and C.A. acknowledge support from the National Science Foundation (Grant No. 2231328).
\end{acknowledgments}

\bibliographystyle{apsrev4-2}
\bibliography{ReferencesBib}

@PREAMBLE{
 "\providecommand{\noopsort}[1]{}" 
 # "\providecommand{\singleletter}[1]{#1}%" 
}

@article{google2020hartree,
  title        = {Hartree–Fock on a superconducting qubit quantum computer},
  author       = {Google AI Quantum and Collaborators and Arute, Frank and Arya, Kunal and Babbush, Ryan and Bacon, Dave and Bardin, Joseph C and Barends, Rami and Boixo, Sergio and Broughton, Michael and Buckley, Bob B and others},
  journal      = {Science},
  volume       = {369},
  number       = {6507},
  pages        = {1084--1089},
  year         = {2020},
  publisher    = {American Association for the Advancement of Science},
  doi          = {10.1126/science.abb9811},
  url          = {https://doi.org/10.1126/science.abb9811}
}

@article{gresch2025guaranteed,
  title        = {Guaranteed efficient energy estimation of quantum many-body Hamiltonians using {ShadowGrouping}},
  author       = {Gresch, Alexander and Kliesch, Martin},
  journal      = {Nature Communications},
  volume       = {16},
  number       = {1},
  pages        = {689},
  year         = {2025},
  publisher    = {Nature Publishing Group UK London},
  doi          = {10.1038/s41467-024-54859-x},
  url          = {https://www.nature.com/articles/s41467-024-54859-x}
}

@article{clinton2024towards,
  title        = {Towards near-term quantum simulation of materials},
  author       = {Clinton, Laura and Cubitt, Toby and Flynn, Brian and Gambetta, Filippo Maria and Klassen, Joel and Montanaro, Ashley and Piddock, Stephen and Santos, Raul A and Sheridan, Evan},
  journal      = {Nature Communications},
  volume       = {15},
  number       = {1},
  pages        = {211},
  year         = {2024},
  publisher    = {Nature Publishing Group UK London},
  doi          = {10.1038/s41467-023-43479-6},
  url          = {https://www.nature.com/articles/s41467-023-43479-6}
}

@article{yoshioka2025krylov,
  title        = {Krylov diagonalization of large many-body Hamiltonians on a quantum processor},
  author       = {Yoshioka, Nobuyuki and Amico, Mirko and Kirby, William and Jurcevic, Petar and Dutt, Arkopal and Fuller, Bryce and Garion, Shelly and Haas, Holger and Hamamura, Ikko and Ivrii, Alexander and others},
  journal      = {Nature Communications},
  volume       = {16},
  number       = {1},
  pages        = {5014},
  year         = {2025},
  publisher    = {Nature Publishing Group UK London},
  doi          = {10.1038/s41467-025-59716-z},
  url          = {https://www.nature.com/articles/s41467-025-59716-z}
}

@article{pelofske2024short,
  title        = {Short-depth {QAOA} circuits and quantum annealing on higher-order {Ising} models},
  author       = {Pelofske, Elijah and B{\"a}rtschi, Andreas and Eidenbenz, Stephan},
  journal      = {npj Quantum Information},
  volume       = {10},
  number       = {1},
  pages        = {30},
  year         = {2024},
  publisher    = {Nature Publishing Group UK London},
  doi          = {10.21203/rs.3.rs-3238348/v1},
  url          = {https://doi.org/10.21203/rs.3.rs-3238348/v1}
}

@article{sciorilli2025towards,
  title        = {Towards large-scale quantum optimization solvers with few qubits},
  author       = {Sciorilli, Marco and Borges, Lucas and Patti, Taylor L and Garc{\'\i}a-Mart{\'\i}n, Diego and Camilo, Giancarlo and Anandkumar, Anima and Aolita, Leandro},
  journal      = {Nature Communications},
  volume       = {16},
  number       = {1},
  pages        = {476},
  year         = {2025},
  publisher    = {Nature Publishing Group UK London},
  doi          = {10.1038/s41467-024-55346-z},
  url          = {https://www.nature.com/articles/s41467-024-55346-z}
}

@article{dunjko2020non,
  title        = {A non-review of quantum machine learning: Trends and explorations},
  author       = {Dunjko, Vedran and Wittek, Peter},
  journal      = {Quantum Views},
  volume       = {4},
  pages        = {32},
  year         = {2020},
  publisher    = {Verein zur F{\"o}rderung des Open Access Publizierens in den Quantenwissenschaften},
  doi          = {10.22331/qv-2020-03-17-32},
  url          = {https://doi.org/10.22331/qv-2020-03-17-32}
}

@article{cho2024machine,
  title        = {Machine learning on quantum experimental data toward solving quantum many-body problems},
  author       = {Cho, Gyungmin and Kim, Dohun},
  journal      = {Nature Communications},
  volume       = {15},
  number       = {1},
  pages        = {7552},
  year         = {2024},
  publisher    = {Nature Publishing Group UK London},
  doi          = {10.1038/s41467-024-51932-3},
  url          = {https://www.nature.com/articles/s41467-024-51932-3}
}

@article{pan2023deep,
  title        = {Deep quantum neural networks on a superconducting processor},
  author       = {Pan, Xiaoxuan and Lu, Zhide and Wang, Weiting and Hua, Ziyue and Xu, Yifang and Li, Weikang and Cai, Weizhou and Li, Xuegang and Wang, Haiyan and Song, Yi-Pu and others},
  journal      = {Nature Communications},
  volume       = {14},
  number       = {1},
  pages        = {4006},
  year         = {2023},
  publisher    = {Nature Publishing Group UK London},
  doi          = {10.1038/s41467-023-39785-8},
  url          = {https://www.nature.com/articles/s41467-023-39785-8}
}

@article{chen2024empowering,
  title        = {Empowering deep neural quantum states through efficient optimization},
  author       = {Chen, Ao and Heyl, Markus},
  journal      = {Nature Physics},
  volume       = {20},
  number       = {9},
  pages        = {1476--1481},
  year         = {2024},
  publisher    = {Nature Publishing Group UK London},
  doi          = {10.1038/s41567-024-02566-1},
  url          = {https://www.nature.com/articles/s41567-024-02566-1}
}

@article{lucas2014ising,
  title        = {Ising formulations of many {NP} problems},
  author       = {Lucas, Andrew},
  journal      = {Frontiers in Physics},
  volume       = {2},
  pages        = {5},
  year         = {2014},
  doi          = {10.3389/fphy.2014.00005},
  url          = {https://www.frontiersin.org/articles/10.3389/fphy.2014.00005/full}
}

@article{farhi2014quantum,
  title        = {A quantum approximate optimization algorithm},
  author       = {Farhi, Edward and Goldstone, Jeffrey and Gutmann, Sam},
  journal      = {arXiv preprint arXiv:1411.4028},
  year         = {2014},
  url          = {https://arxiv.org/abs/1411.4028}
}

@article{mohseni2022ising,
  title        = {Ising machines as hardware solvers of combinatorial optimization problems},
  author       = {Mohseni, Naeimeh and McMahon, Peter L and Byrnes, Tim},
  journal      = {Nature Reviews Physics},
  volume       = {4},
  number       = {6},
  pages        = {363--379},
  year         = {2022},
  publisher    = {Nature Publishing Group UK London},
  doi          = {10.1038/s42254-022-00440-8},
  url          = {https://www.nature.com/articles/s42254-022-00440-8}
}

@article{jiang2023efficient,
  title        = {Efficient combinatorial optimization by quantum-inspired parallel annealing in analogue memristor crossbar},
  author       = {Jiang, Mingrui and Shan, Keyi and He, Chengping and Li, Can},
  journal      = {Nature Communications},
  volume       = {14},
  number       = {1},
  pages        = {5927},
  year         = {2023},
  publisher    = {Nature Publishing Group UK London},
  doi          = {10.1038/s41467-023-41647-2},
  url          = {https://www.nature.com/articles/s41467-023-41647-2}
}

@article{bybee2023efficient,
  title        = {Efficient optimization with higher-order {Ising} machines},
  author       = {Bybee, Connor and Kleyko, Denis and Nikonov, Dmitri E and Khosrowshahi, Amir and Olshausen, Bruno A and Sommer, Friedrich T},
  journal      = {Nature Communications},
  volume       = {14},
  number       = {1},
  pages        = {6033},
  year         = {2023},
  publisher    = {Nature Publishing Group UK London},
  doi          = {10.1038/s41467-023-41214-9},
  url          = {https://www.nature.com/articles/s41467-023-41214-9}
}

@article{zhang2024review,
  title        = {A review of {Ising} machines implemented in conventional and emerging technologies},
  author       = {Zhang, Tingting and Tao, Qichao and Liu, Bailiang and Grimaldi, Andrea and Raimondo, Eleonora and Jim{\'e}nez, Manuel and Avedillo, Mar{\'\i}a Jos{\'e} and Nu{\~n}ez, Juan and Linares-Barranco, Bernab{\'e} and Serrano-Gotarredona, Teresa and others},
  journal      = {IEEE Transactions on Nanotechnology},
  volume       = {23},
  pages        = {704--717},
  year         = {2024},
  publisher    = {IEEE},
  doi          = {10.1109/TNANO.2024.3457533},
  url          = {https://dl.acm.org/doi/10.1109/TNANO.2024.3457533}
}

@article{si2024energy,
  title        = {Energy-efficient superparamagnetic {Ising} machine and its application to traveling salesman problems},
  author       = {Si, Jia and Yang, Shuhan and Cen, Yunuo and Chen, Jiaer and Huang, Yingna and Yao, Zhaoyang and Kim, Dong-Jun and Cai, Kaiming and Yoo, Jerald and Fong, Xuanyao and others},
  journal      = {Nature Communications},
  volume       = {15},
  number       = {1},
  pages        = {3457},
  year         = {2024},
  publisher    = {Nature Publishing Group UK London},
  doi          = {10.1038/s41467-024-47818-z},
  url          = {https://www.nature.com/articles/s41467-024-47818-z}
}

@article{ushijima2017graph,
  title        = {Graph partitioning using quantum annealing on the {D}-{Wave} system},
  author       = {Ushijima-Mwesigwa, Hayato and Negre, Christian F. A. and Mniszewski, Susan M},
  journal      = {arXiv preprint arXiv:1705.03082},
  year         = {2017},
  url          = {https://arxiv.org/abs/1705.03082}
}

@article{pelofske2021decomposition,
  title        = {Decomposition algorithms for solving {NP}-hard problems on a quantum annealer},
  author       = {Pelofske, Elijah and Hahn, Georg and Djidjev, Hristo},
  journal      = {Journal of Signal Processing Systems},
  volume       = {93},
  number       = {4},
  pages        = {405--420},
  year         = {2021},
  publisher    = {Springer},
  url          = {https://link.springer.com/article/10.1007/s11265-020-01550-1}
}

@article{bozejko2025optimal,
  title        = {Optimal solving of a scheduling problem using quantum annealing metaheuristics on the {D}-{Wave} quantum solver},
  author       = {Bo{\.z}ejko, Wojciech and Klempous, Ryszard and Pempera, Jaros{\l}aw and Rozenblit, Jerzy and Smutnicki, Czes{\l}aw and Uchro{\'n}ski, Mariusz},
  journal      = {IEEE Transactions on Systems, Man, and Cybernetics: Systems},
  volume       = {55},
  number       = {1},
  pages        = {196--208},
  year         = {2025},
  doi          = {10.1109/TSMC.2024.3458873},
  publisher    = {IEEE},
  url          = {https://doi.org/10.1109/TSMC.2024.3458873}
}

@article{stollenwerk2019quantum,
  title        = {Quantum annealing applied to de-conflicting optimal trajectories for air traffic management},
  author       = {Stollenwerk, Tobias and O’Gorman, Bryan and Venturelli, Davide and Mandra, Salvatore and Rodionova, Olga and Ng, Hokkwan and Sridhar, Banavar and Rieffel, Eleanor Gilbert and Biswas, Rupak},
  journal      = {IEEE Transactions on Intelligent Transportation Systems},
  volume       = {21},
  number       = {1},
  pages        = {285--297},
  year         = {2019},
  publisher    = {IEEE},
  url          = {https://ieeexplore.ieee.org/abstract/document/8643733}
}

@article{fu2025solving,
  title        = {Solving flexible job-shop scheduling problems based on quantum computing},
  author       = {Fu, Kaihan and Liu, Jianjun and Chen, Miao and Zhang, Huiying},
  journal      = {Entropy},
  volume       = {27},
  number       = {2},
  pages        = {189},
  year         = {2025},
  doi          = {10.3390/e27020189},
  url          = {https://doi.org/10.3390/e27020189},
  publisher    = {MDPI}
}

@article{ajagekar2022hybrid,
  title        = {Hybrid classical--quantum optimization techniques for solving mixed-integer programming problems in production scheduling},
  author       = {Ajagekar, Akshay and Al Hamoud, Kumail and You, Fengqi},
  journal      = {IEEE Transactions on Quantum Engineering},
  volume       = {3},
  pages        = {1--16},
  year         = {2022},
  doi          = {10.1109/TQE.2022.3187367},
  publisher    = {IEEE},
  url          = {https://doi.org/10.1109/TQE.2022.3187367}
}

@article{ohyama2023resource,
  title        = {Resource allocation optimization by quantum computing for shared use of standalone {IRS}},
  author       = {Ohyama, Takahiro and Kawamoto, Yuichi and Kato, Nei},
  journal      = {IEEE Transactions on Emerging Topics in Computing},
  volume       = {11},
  number       = {4},
  pages        = {950--961},
  year         = {2023},
  doi          = {10.1109/TETC.2023.3292355},
  publisher    = {IEEE},
  url          = {https://doi.org/10.1109/TETC.2023.3292355}
}

@article{butt2024quantum,
  title        = {Quantum-inspired resource optimization for {6G} networks: A survey},
  author       = {Butt, Muhammad Omair and Waheed, Nazar and Duong, Trung Q. and Ejaz, Waleed},
  journal      = {IEEE Communications Surveys \& Tutorials},
  year         = {2024},
  pages        = {1--1},
  doi          = {10.1109/COMST.2024.3519865},
  publisher    = {IEEE},
  note         = {Early Access},
  url          = {https://doi.org/10.1109/COMST.2024.3519865}
}

@article{cerezo2021variational,
  title        = {Variational quantum algorithms},
  author       = {Cerezo, Marco and Arrasmith, Andrew and Babbush, Ryan and Benjamin, Simon C and Endo, Suguru and Fujii, Keisuke and McClean, Jarrod R and Mitarai, Kosuke and Yuan, Xiao and Cincio, Lukasz and others},
  journal      = {Nature Reviews Physics},
  volume       = {3},
  number       = {9},
  pages        = {625--644},
  year         = {2021},
  publisher    = {Nature Publishing Group UK London},
  doi          = {10.1038/s42254-021-00348-9},
  url          = {https://doi.org/10.1038/s42254-021-00348-9}
}

@article{kandala2017hardware,
  title        = {Hardware-efficient variational quantum eigensolver for small molecules and quantum magnets},
  author       = {Kandala, Abhinav and Mezzacapo, Antonio and Temme, Kristan and Takita, Maika and Brink, Markus and Chow, Jerry M and Gambetta, Jay M},
  journal      = {Nature},
  volume       = {549},
  number       = {7671},
  pages        = {242--246},
  year         = {2017},
  publisher    = {Nature Publishing Group},
  doi          = {10.1038/nature23879},
  url          = {https://doi.org/10.1038/nature23879}
}

@article{parrish2019quantum,
  title        = {Quantum computation of electronic transitions using a variational quantum eigensolver},
  author       = {Parrish, Robert M and Hohenstein, Edward G and McMahon, Peter L and Mart{\'\i}nez, Todd J},
  journal      = {Physical Review Letters},
  volume       = {122},
  number       = {23},
  pages        = {230401},
  year         = {2019},
  publisher    = {APS},
  doi          = {10.1103/PhysRevLett.122.230401},
  url          = {https://doi.org/10.1103/PhysRevLett.122.230401}
}

@article{liu2019variational,
  title        = {Variational quantum eigensolver with fewer qubits},
  author       = {Liu, Jin-Guo and Zhang, Yi-Hong and Wan, Yuan and Wang, Lei},
  journal      = {Physical Review Research},
  volume       = {1},
  number       = {2},
  pages        = {023025},
  year         = {2019},
  publisher    = {APS},
  doi          = {10.1103/PhysRevResearch.1.023025},
  url          = {https://doi.org/10.1103/PhysRevResearch.1.023025}
}

@article{barison2022variational,
  title        = {Variational dynamics as a ground-state problem on a quantum computer},
  author       = {Barison, Stefano and Vicentini, Filippo and Cirac, Ignacio and Carleo, Giuseppe},
  journal      = {Physical Review Research},
  volume       = {4},
  number       = {4},
  pages        = {043161},
  year         = {2022},
  publisher    = {APS},
  doi          = {10.1103/PhysRevResearch.4.043161},
  url          = {https://doi.org/10.1103/PhysRevResearch.4.043161}
}

@article{fujii2022deep,
  title        = {Deep variational quantum eigensolver: A divide-and-conquer method for solving a larger problem with smaller size quantum computers},
  author       = {Fujii, Keisuke and Mizuta, Kaoru and Ueda, Hiroshi and Mitarai, Kosuke and Mizukami, Wataru and Nakagawa, Yuya O},
  journal      = {PRX Quantum},
  volume       = {3},
  number       = {1},
  pages        = {010346},
  year         = {2022},
  publisher    = {APS},
  doi          = {10.1103/PRXQuantum.3.010346},
  url          = {https://doi.org/10.1103/PRXQuantum.3.010346}
}

@article{wang2024entanglement,
  title        = {Entanglement-variational hardware-efficient ansatz for eigensolvers},
  author       = {Wang, Xin and Qi, Bo and Wang, Yabo and Dong, Daoyi},
  journal      = {Physical Review Applied},
  volume       = {21},
  number       = {3},
  pages        = {034059},
  year         = {2024},
  publisher    = {APS},
  doi          = {10.1103/PhysRevApplied.21.034059},
  url          = {https://doi.org/10.1103/PhysRevApplied.21.034059}
}

@article{mcclean2016theory,
  title        = {The theory of variational hybrid quantum-classical algorithms},
  author       = {McClean, Jarrod R and Romero, Jonathan and Babbush, Ryan and Aspuru-Guzik, Al{\'a}n},
  journal      = {New Journal of Physics},
  volume       = {18},
  number       = {2},
  pages        = {023023},
  year         = {2016},
  publisher    = {IOP Publishing},
  doi          = {10.1088/1367-2630/18/2/023023},
  url          = {https://doi.org/10.1088/1367-2630/18/2/023023}
}

@article{bittel2021training,
  title        = {Training variational quantum algorithms is {NP}-hard},
  author       = {Bittel, Lennart and Kliesch, Martin},
  journal      = {Physical Review Letters},
  volume       = {127},
  number       = {12},
  pages        = {120502},
  year         = {2021},
  publisher    = {APS},
  doi          = {10.1103/PhysRevLett.127.120502},
  url          = {https://doi.org/10.1103/PhysRevLett.127.120502}
}

@article{grimsley2019adaptive,
  title        = {An adaptive variational algorithm for exact molecular simulations on a quantum computer},
  author       = {Grimsley, Harper R and Economou, Sophia E and Barnes, Edwin and Mayhall, Nicholas J},
  journal      = {Nature Communications},
  volume       = {10},
  number       = {1},
  pages        = {3007},
  year         = {2019},
  publisher    = {Nature Publishing Group UK London},
  doi          = {10.1038/s41467-019-10988-2},
  url          = {https://doi.org/10.1038/s41467-019-10988-2}
}

@article{tang2021qubit,
  title        = {{qubit-adapt-vqe}: An adaptive algorithm for constructing hardware-efficient ans{\"a}tze on a quantum processor},
  author       = {Tang, Ho Lun and Shkolnikov, V. O. and Barron, George S. and Grimsley, Harper R. and Mayhall, Nicholas J. and Barnes, Edwin and Economou, Sophia E},
  journal      = {PRX Quantum},
  volume       = {2},
  number       = {2},
  pages        = {020310},
  year         = {2021},
  publisher    = {APS},
  doi          = {10.1103/PRXQuantum.2.020310},
  url          = {https://doi.org/10.1103/PRXQuantum.2.020310}
}

@article{schulte2010gradient,
  title        = {Gradient flows for optimization in quantum information and quantum dynamics: Foundations and applications},
  author       = {Schulte-Herbr{\"u}ggen, Thomas and Glaser, Steffen J and Dirr, Gunther and Helmke, Uwe},
  journal      = {Reviews in Mathematical Physics},
  volume       = {22},
  number       = {06},
  pages        = {597--667},
  year         = {2010},
  publisher    = {World Scientific},
  doi          = {10.1142/S0129055X10004053},
  url          = {https://doi.org/10.1142/S0129055X10004053}
}

@article{panageas2019first,
  title        = {First-order methods almost always avoid saddle points: The case of vanishing step-sizes},
  author       = {Panageas, Ioannis and Piliouras, Georgios and Wang, Xiao},
  journal      = {Advances in Neural Information Processing Systems},
  volume       = {32},
  year         = {2019},
  doi          = {10.48550/arXiv.1710.07406},
  url          = {https://doi.org/10.48550/arXiv.1710.07406}
}

@article{malvetti2024randomized,
  title        = {Randomized gradient descents on {Riemannian} manifolds: Almost sure convergence to global minima in and beyond quantum optimization},
  author       = {Malvetti, Emanuel and Arenz, Christian and Dirr, Gunther and Schulte-Herbr{\"u}ggen, Thomas},
  journal      = {arXiv preprint arXiv:2405.12039},
  year         = {2024},
  doi          = {10.48550/arXiv.2405.12039},
  url          = {https://doi.org/10.48550/arXiv.2405.12039}
}

@article{mcmahon2025equating,
  title        = {Equating quantum imaginary time evolution, {Riemannian} gradient flows, and stochastic implementations},
  author       = {McMahon, Nathan A and Pervez, Mahum and Arenz, Christian},
  journal      = {arXiv preprint arXiv:2504.06123},
  year         = {2025},
  url          = {https://arxiv.org/abs/2504.06123}
}

@article{magann2023randomized,
  title        = {Randomized adaptive quantum state preparation},
  author       = {Magann, Alicia B and Economou, Sophia E and Arenz, Christian},
  journal      = {Physical Review Research},
  volume       = {5},
  number       = {3},
  pages        = {033227},
  year         = {2023},
  publisher    = {APS},
  doi          = {10.1103/PhysRevResearch.5.033227},
  url          = {https://doi.org/10.1103/PhysRevResearch.5.033227}
}

@article{wiersema2023optimizing,
  title        = {Optimizing quantum circuits with {Riemannian} gradient flow},
  author       = {Wiersema, Roeland and Killoran, Nathan},
  journal      = {Physical Review A},
  volume       = {107},
  number       = {6},
  pages        = {062421},
  year         = {2023},
  publisher    = {APS},
  doi          = {10.1103/PhysRevA.107.062421},
  url          = {https://doi.org/10.1103/PhysRevA.107.062421}
}

@article{suzuki1985decomposition,
  title        = {Decomposition formulas of exponential operators and {Lie} exponentials with some applications to quantum mechanics and statistical physics},
  author       = {Suzuki, Masuo},
  journal      = {Journal of Mathematical Physics},
  volume       = {26},
  number       = {4},
  pages        = {601--612},
  year         = {1985},
  publisher    = {AIP Publishing},
  url          = {https://doi.org/10.1063/1.526596}
}

@article{campbell2019random,
  title        = {Random compiler for fast {Hamiltonian} simulation},
  author       = {Campbell, Earl},
  journal      = {Physical Review Letters},
  volume       = {123},
  number       = {7},
  pages        = {070503},
  year         = {2019},
  publisher    = {APS},
  url          = {https://doi.org/10.1103/PhysRevLett.123.070503}
}

@article{rabitz2004quantum,
  title        = {Quantum optimally controlled transition landscapes},
  author       = {Rabitz, Herschel A and Hsieh, Michael M and Rosenthal, Carey M},
  journal      = {Science},
  volume       = {303},
  number       = {5666},
  pages        = {1998--2001},
  year         = {2004},
  publisher    = {American Association for the Advancement of Science},
  url          = {https://www.science.org/doi/full/10.1126/science.1093649}
}

@article{hsieh2009topology,
  title        = {Topology of the quantum control landscape for observables},
  author       = {Hsieh, Michael and Wu, Rebing and Rabitz, Herschel},
  journal      = {The Journal of Chemical Physics},
  volume       = {130},
  number       = {10},
  year         = {2009},
  publisher    = {AIP Publishing},
  url          = {https://pubs.aip.org/aip/jcp/article/130/10/104109/906281}
}

@article{bennett1997strengths,
  title        = {Strengths and weaknesses of quantum computing},
  author       = {Bennett, Charles H and Bernstein, Ethan and Brassard, Gilles and Vazirani, Umesh},
  journal      = {SIAM Journal on Computing},
  volume       = {26},
  number       = {5},
  pages        = {1510--1523},
  year         = {1997},
  publisher    = {SIAM},
  url={https://doi.org/10.1137/S0097539796300933}
}

@book{bondy1976graph,
  title        = {Graph theory with applications},
  author       = {Bondy, John Adrian and Murty, Uppaluri Siva Ramachandra and others},
  volume       = {290},
  year         = {1976},
  publisher    = {Macmillan London},
  url          = {https://dl.acm.org/doi/abs/10.5555/1097029}
}

@book{west2001introduction,
  title        = {Introduction to graph theory},
  author       = {West, Douglas B},
  year         = {2001},
  publisher    = {Prentice Hall},
  url          = {https://books.google.com/books/about/Introduction_to_Graph_Theory.html?id=TuvuAAAAMAAJ}
}

@article{Lloyd1996,
  title        = {Universal quantum simulators},
  author       = {Lloyd, Seth},
  journal      = {Science},
  year         = {1996},
  volume       = {273},
  pages        = {1073--1078},
  url          = {https://www.science.org/doi/10.1126/science.273.5278.1073}
}

@article{Berry2015Taylor,
  title        = {Simulating {Hamiltonian} dynamics with a truncated {Taylor} series},
  author       = {Berry, Dominic W. and Childs, Andrew M. and Cleve, Richard and Kothari, Robin and Somma, Rolando D.},
  journal      = {Physical Review Letters},
  year         = {2015},
  volume       = {114},
  pages        = {090502},
  url          = {https://doi.org/10.1103/PhysRevLett.114.090502}
}

@article{LowChuang2019,
  title        = {Hamiltonian simulation by qubitization},
  author       = {Low, Guang Hao and Chuang, Isaac L.},
  journal      = {Quantum},
  year         = {2019},
  volume       = {3},
  pages        = {163},
  url          = {https://doi.org/10.22331/q-2019-07-12-163}
}

@article{Childs2019PF,
  title        = {A theory of {Trotter} error with commutator scaling},
  author       = {Childs, Andrew M. and Su, Yuan and Tran, Minh C. and Wiebe, Nathan and Zhu, Shuchen},
  journal      = {Proceedings of the National Academy of Sciences of the United States of America},
  year         = {2019},
  volume       = {116},
  number       = {42},
  pages        = {20726--20732},
  url          = {https://doi.org/10.1103/PhysRevX.11.011020}
}

@book{helmke2012optimization,
  title        = {Optimization and dynamical systems},
  author       = {Helmke, Uwe and Moore, John B},
  year         = {2012},
  publisher    = {Springer Science \& Business Media},
  doi          = {10.1137/1038093},
  url          = {https://doi.org/10.1137/1038093}
}

@book{absil2008optimization,
  title        = {Optimization algorithms on matrix manifolds},
  author       = {Absil, P-A and Mahony, Robert and Sepulchre, Rodolphe},
  year         = {2008},
  publisher    = {Princeton University Press},
  doi          = {10.1515/9781400830244},
  url          = {https://doi.org/10.1515/9781400830244}
}

@article{schuld2019evaluating,
  title        = {Evaluating analytic gradients on quantum hardware},
  author       = {Schuld, Maria and Bergholm, Ville and Gogolin, Christian and Izaac, Josh and Killoran, Nathan},
  journal      = {Physical Review A},
  volume       = {99},
  number       = {3},
  pages        = {032331},
  year         = {2019},
  publisher    = {APS},
  doi          = {10.1103/PhysRevA.99.032331},
  url          = {https://doi.org/10.1103/PhysRevA.99.032331}
}

@article{mitarai2018quantum,
  title        = {Quantum circuit learning},
  author       = {Mitarai, Kosuke and Negoro, Makoto and Kitagawa, Masahiro and Fujii, Keisuke},
  journal      = {Physical Review A},
  volume       = {98},
  number       = {3},
  pages        = {032309},
  year         = {2018},
  publisher    = {APS},
  doi          = {10.1103/PhysRevA.98.032309},
  url          = {https://doi.org/10.1103/PhysRevA.98.032309}
}

@article{wiedmann2025convergence,
  title={On the convergence of the variational quantum eigensolver and quantum optimal control},
  author={Wiedmann, Marco and Burgarth, Daniel and Dirr, Gunther and Schulte-Herbr{\"u}ggen, Thomas and Malvetti, Emanuel and Arenz, Christian},
  journal={arXiv preprint arXiv:2509.05295},
  year={2025},
  url={https://doi.org/10.48550/arXiv.2509.05295}
}

@article{holmes2022connecting,
  title={Connecting ansatz expressibility to gradient magnitudes and barren plateaus},
  author={Holmes, Zo{\"e} and Sharma, Kunal and Cerezo, Marco and Coles, Patrick J},
  journal={PRX quantum},
  volume={3},
  number={1},
  pages={010313},
  year={2022},
  publisher={APS},
  url={https://doi.org/10.1103/PRXQuantum.3.010313}
}

@article{magann2022feedback,
  title={Feedback-based quantum optimization},
  author={Magann, Alicia B and Rudinger, Kenneth M and Grace, Matthew D and Sarovar, Mohan},
  journal={Physical Review Letters},
  volume={129},
  number={25},
  pages={250502},
  year={2022},
  publisher={APS},
  url={https://doi.org/10.1103/PhysRevLett.129.250502}
}

@article{gluza2024double,
  title={Double-bracket quantum algorithms for quantum imaginary-time evolution},
  author={Gluza, Marek and Son, Jeongrak and Tiang, Bi Hong and Zander, Ren{\'e} and Seidel, Raphael and Suzuki, Yudai and Holmes, Zo{\"e} and Ng, Nelly HY},
  journal={arXiv preprint arXiv:2412.04554},
  year={2024},
  url={https://doi.org/10.48550/arXiv.2412.04554}
}

@article{x8g1-7h1k,
  title = {Nonvariational ADAPT algorithm for quantum simulations},
  author = {Tang, Ho Lun and Chen, Yanzhu and Biswas, Prakriti and Magann, Alicia B. and Arenz, Christian and Economou, Sophia E.},
  journal = {Phys. Rev. Res.},
  volume = {7},
  issue = {2},
  pages = {023275},
  numpages = {15},
  year = {2025},
  month = {Jun},
  publisher = {American Physical Society},
  doi = {10.1103/x8g1-7h1k},
  url = {https://link.aps.org/doi/10.1103/x8g1-7h1k}
}

@article{Sim2019Expressibility,
  title   = {Expressibility and Entangling Capability of Parameterized Quantum Circuits for Hybrid Quantum-Classical Algorithms},
  author  = {Sim, Sukin and Johnson, Peter D. and Aspuru-Guzik, Al{\'a}n},
  journal = {Adv. Quantum Technol.},
  volume  = {2},
  number  = {12},
  pages   = {1900070},
  year    = {2019},
  doi     = {10.1002/qute.201900070},
  url     = {https://doi.org/10.1002/qute.201900070}
}

@inproceedings{zander2025role,
  title={Role of Riemannian geometry in double-bracket quantum imaginary-time evolution},
  author={Zander, Ren{\'e} and Seidel, Raphael and Xiaoyue, Li and Gluza, Marek},
  booktitle={International Conference on Geometric Science of Information},
  pages={105--114},
  year={2025},
  organization={Springer},
  url={https://link.springer.com/chapter/10.1007/978-3-032-03924-8_11}
}

@article{alghadeer2025double,
  title={Double-bracket algorithmic cooling},
  author={Alghadeer, Mohammed and Giang, Khanh Uyen and Cao, Shuxiang and Fasciati, Simone D and Piscitelli, Michele and Ng, Nelly and Leek, Peter J and Gluza, Marek and Bakr, Mustafa},
  journal={arXiv preprint arXiv:2510.00302},
  year={2025},
  url={
https://doi.org/10.48550/arXiv.2510.00302}
}

@article{PhysRevLett.134.180602,
  title = {Quantum Dynamic Programming},
  author = {Son, Jeongrak and Gluza, Marek and Takagi, Ryuji and Ng, Nelly H. Y.},
  journal = {Phys. Rev. Lett.},
  volume = {134},
  issue = {18},
  pages = {180602},
  numpages = {8},
  year = {2025},
  month = {May},
  publisher = {American Physical Society},
  doi = {10.1103/PhysRevLett.134.180602},
  url = {https://link.aps.org/doi/10.1103/PhysRevLett.134.180602}
}

@article{robbiati2024double,
  title={Double-bracket quantum algorithms for high-fidelity ground state preparation},
  author={Robbiati, Matteo and Pedicillo, Edoardo and Pasquale, Andrea and Li, Xiaoyue and Kiss, Oriel and Wright, Andrew and Farias, Renato and Giang, Khanh Uyen and Son, Jeongrak and Kn{\"o}rzer, Johannes and others},
  journal={arXiv preprint arXiv:2408.03987},
  year={2024},
  url={
https://doi.org/10.48550/arXiv.2408.03987}
}

@article{suzuki2025grover,
  title={Grover's algorithm is an approximation of imaginary-time evolution},
  author={Suzuki, Yudai and Gluza, Marek and Son, Jeongrak and Tiang, Bi Hong and Ng, Nelly HY and Holmes, Zo{\"e}},
  journal={arXiv preprint arXiv:2507.15065},
  year={2025},
  url={
https://doi.org/10.48550/arXiv.2507.15065}
}

@article{suzuki2025double,
  title={Double-bracket algorithm for quantum signal processing without post-selection},
  author={Suzuki, Yudai and Tiang, Bi Hong and Son, Jeongrak and Ng, Nelly HY and Holmes, Zo{\"e} and Gluza, Marek},
  journal={arXiv preprint arXiv:2504.01077},
  year={2025},
  url={
https://doi.org/10.48550/arXiv.2504.01077}
}

@article{xiaoyue2024strategies,
  title={Strategies for optimizing double-bracket quantum algorithms},
  author={Xiaoyue, Li and Robbiati, Matteo and Pasquale, Andrea and Pedicillo, Edoardo and Wright, Andrew and Carrazza, Stefano and Gluza, Marek},
  journal={arXiv preprint arXiv:2408.07431},
  year={2024},
  url={
https://doi.org/10.48550/arXiv.2408.07431}
}

\end{document}